\documentclass[a4paper]{article}
\usepackage[utf8]{inputenc}
\usepackage{authblk}
\usepackage{setspace}
\usepackage[margin=1.5cm]{geometry}
\usepackage{graphicx}
\graphicspath{ {./Figures/} }
\usepackage{subcaption}
\usepackage{amsmath}

\usepackage{siunitx}
\usepackage{float}
\usepackage{caption}
\usepackage{subcaption}
\usepackage{upgreek}
\usepackage{enumitem}
\usepackage{graphicx}
\usepackage [utf8]{inputenc}
\usepackage[T1]{fontenc}
\usepackage{epsfig}
\setlist{nosep} % or \setlist{noitemsep} to leave space around whole list

\usepackage{fancyhdr}
\title{\large \textbf{ Few-cycle pulse generation by double-stage hybrid multi-pass multi-plate nonlinear pulse compression}}

\author[1,2]{\normalsize Anne-Lise Viotti}
\author[1]{Chen Li}
\author[3]{Gunnar Arisholm}
\author[1]{Lutz Winkelmann}
\author[1]{Ingmar Hartl}
\author[1,4,5]{Christoph M. Heyl}
\author[1,*]{Marcus Seidel}

\affil[1]{Deutsches Elektronen-Synchrotron DESY, Notkestr. 85, 22607 Hamburg, Germany}
\affil[2]{Department of Physics, Lund University, P.O. Box 118, SE-221 00 Lund, Sweden}
\affil[3]{FFI (Norwegian Defence Research Establishment), P. O. Box 25, NO-2027 Kjeller, Norway}
\affil[4]{Helmholtz-Institute Jena, Fröbelstieg 3,07743 Jena, Germany}
\affil[5]{GSI Helmholtzzentrum für Schwerionenforschnung GmbH, Planckstrasse 1, 64291 Darmstadt, Germany}

\affil[*]{\textit{Corresponding author: marcus.seidel@desy.de}}

\date{}

\newcommand{\muJ}{\,$\mu$J}
\newcommand{\Fig}[1]{Fig.~\ref{#1}}

\begin{document}

\twocolumn[
\maketitle

%%%% Body %%%%

\textbf{{Abstract}\\}
	Few-cycle pulses present an essential tool to track ultrafast dynamics in matter and drive strong field effects. To address photon-hungry applications, high average power lasers are used which, however, cannot directly provide sub-100\,fs pulse durations. Post-compression of laser pulses by spectral broadening and dispersion compensation is the most efficient method to overcome this limitation. Here, we demonstrate a notably compact setup which turns a 0.1\,GW peak power, picosecond laser into a 2.9\,GW peak power, 8.2\,fs source. The 120-fold pulse duration shortening is accomplished in a two-stage hybrid multi-pass, multi-plate compression setup. To our knowledge, neither shorter pulses, nor higher peak powers have been reported to-date from bulk multi-pass cells alone, manifesting the power of the hybrid approach. It puts, for instance, compact, cost-efficient and high repetition rate attosecond sources within reach.\\]

%%%% INTRODUCTION %%%%
Few-cycle pulses have pushed the frontiers of nonlinear optics far beyond the perturbative regime. The (temporary) detachment of weakly bound electrons from the nuclei by strong fields leads to the creation of large electric dipole moments \cite{brabec_intense_2000}. The atomic polarization is switched by few-cycle pulses on sub-femtosecond timescales without prior distortions of the interacting matter \cite{brabec_intense_2000}. Many unique applications emerged, most prominent, the generation of coherent extreme ultraviolet or X-ray radiation and its temporal confinement to attosecond durations \cite{orfanos_attosecond_2019}. This, in turn, enabled tracking of ionization dynamics and performing electron microscopy with highest temporal and spatial resolution \cite{ciappina_attosecond_2017,mikaelsson_high-repetition_2021}. Beyond that, few-cycle pulses prospectively enable PHz bandwidth signal processing in semiconductors, dielectrics and novel quantum materials \cite{kruchinin_colloquium_2018,jimenez-galan_sub-cycle_2021}.
Initial few-cycle sources relied on broadband laser gain media that are difficult to scale in average power \cite{brabec_intense_2000}. However, high pulse repetition rates are important to achieve good signal-to-noise ratios despite the low efficiencies of extremely nonlinear processes or limitations caused by Coulomb interactions after ionization \cite{mikaelsson_high-repetition_2021}. The advancement of ultrafast lasers in the past years to substantially higher average powers \cite{zuo_highpower_2022}, has allowed to overcome the repetition rate short-coming of few-cycle sources, but has also imposed the challenge to reduce the inherent pulse durations of power-scalable lasers from hundreds or thousands of femtoseconds to the sub-10\,fs regime. One approach to accomplish this is optical parametric chirped pulse amplification \cite{furch_high-power-laser_2022}. It provides wavelength tunability and excellent pulse contrast but is a relatively inefficient, complex method. Alternatively, nonlinear spectral broadening and pulse post-compression present a direct, cost-efficient path to the few-cycle regime \cite{nagy_high-energy_2021}. In particular, the multi-pass cell (MPC) spectral broadening technique has combined large pulse compression factors, i.e. the input to output pulse duration ratios, and high power efficiencies in an outstanding manner \cite{schulte_nonlinear_2016,viotti_multi-pass_2022,hanna_MPC_rev_2021}. Recently, several few-cycle pulse generation schemes by means of MPCs have been reported \cite{balla_postcompression_2020,muller_multipass_2021,rueda_8_2021,daniault_single-stage_2021,hadrich_carrier-envelope_2022}. However, all experiments were based on gas-filled MPCs which require hundreds of $\mu$J of pulse energies as well as a chamber that needs to be evacuated and refilled with up to several bars of nonlinear gas. In contrast, bulk material based few-cycle or even single-cycle pulse generation was demonstrated in the past years by the multiple plate continuum approach \cite{lu_generation_2014,lu_greater_2019,seo_high-contrast_2020}. We have recently shown that combining the multiple plate and the bulk MPC techniques can clearly overcome the compression factors that are achievable by the methods alone in a single stage \cite{seidel_factor_2022,seidel_ultrafast_2022}. Here, we apply this novel hybrid approach to demonstrate more than hundred times duration reduction of powerful ultrashort pulses, that is from the picosecond regime to 8.2\,fs FWHM duration. Moreover, we report the first bulk-based MPC that delivers sub-10\,fs pulses with multi-GW peak powers. \par

%%%% RESULTS %%%%%%%%%%%%%%%%%%%%%%%%%%%%%%%%%%%%%%%%%%%%%%%%%%%%%%%%%%%%%%%%%%%%%%%%%%%%%%%%%%%%%%%%%%%%%%%%%%%

%% FIG 1 %%%%%%%%%%%%%%%%%%%%%%%%%%%%%%%%%%%%%%%%%%%%%%%%%%%%%%%%%%%%%%%%%%%%%%%%%%%%%%%%%%%%%%%%%%%%%%%%%
\begin{figure}[t!]
\begin{minipage}[t]{0.6\linewidth}
	\vspace*{\fill}
	\includegraphics[width=\linewidth]{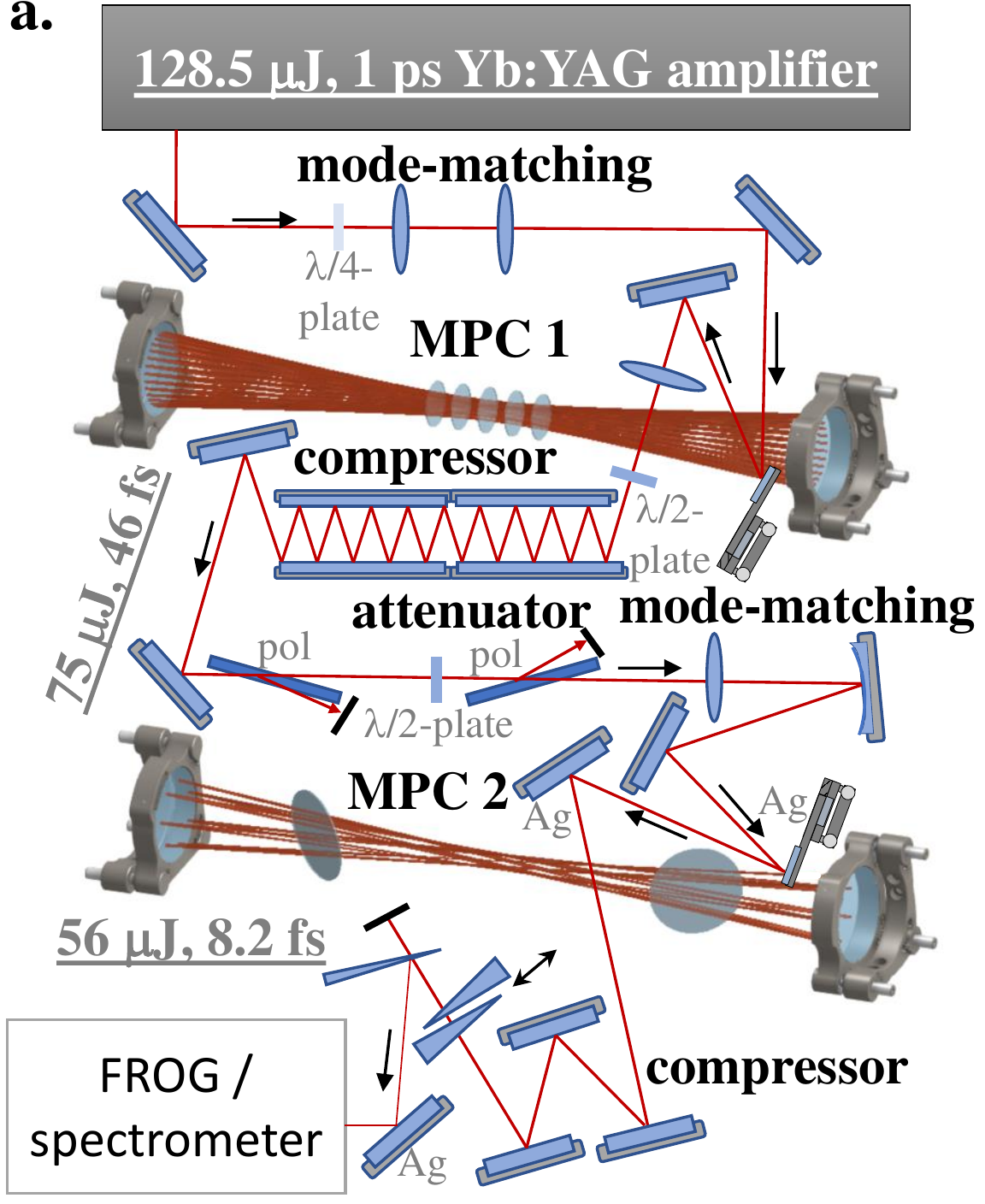}
\end{minipage}
\begin{minipage}[t]{0.39\linewidth}
	\vspace*{-.5mm}
	\includegraphics[width=\linewidth]{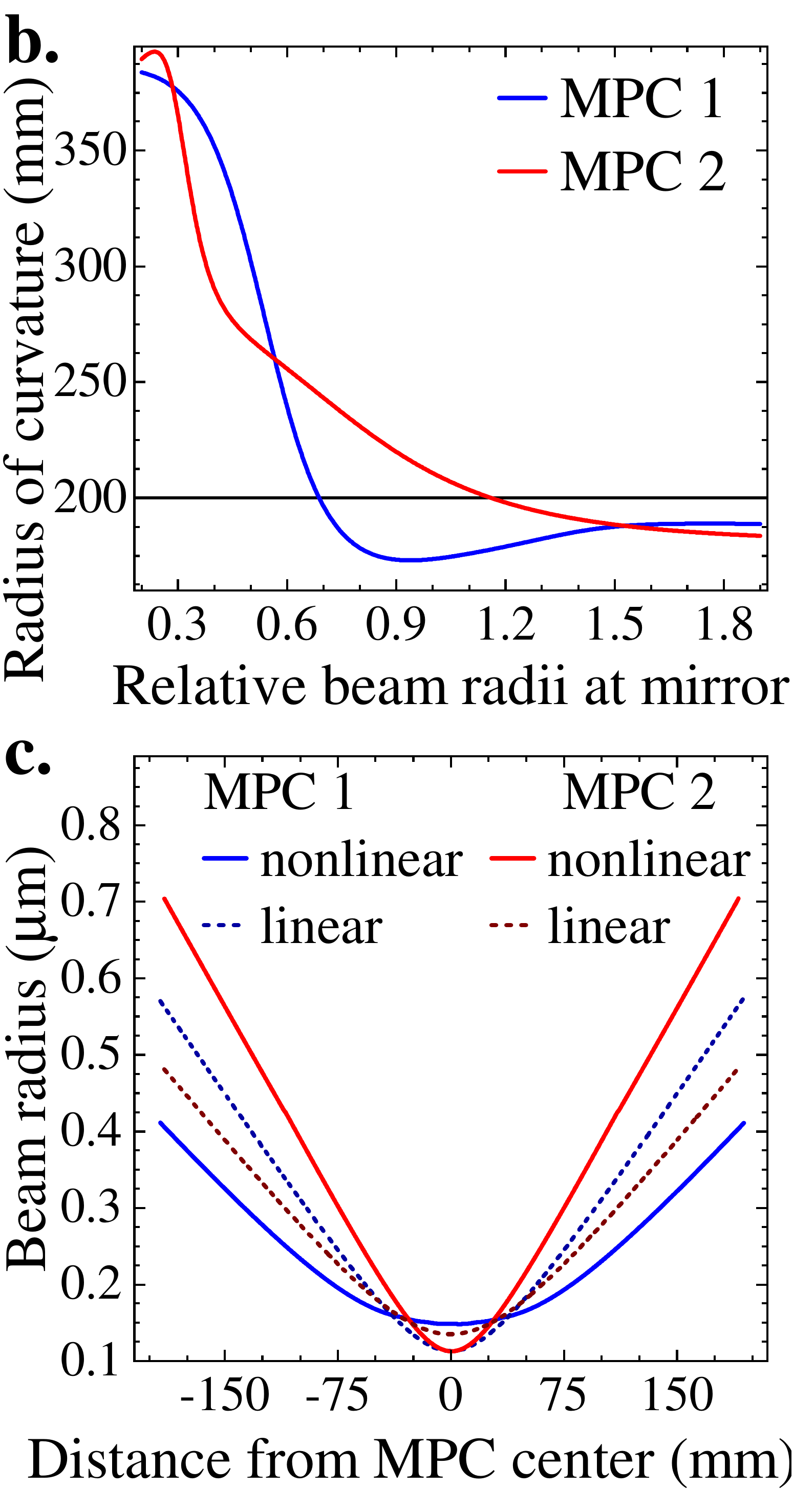}
\end{minipage}
\caption{\textbf{a.} Two-stage pulse compression setup. Both MPC mirror pairs are separated by circa 38\,cm. The compressor and MPC~2 mirrors were chirped. The silver mirrors are denoted by Ag. All other mirrors were quarter-wave stacks. Thin-film polarizers (pol) were used. \textbf{b.} A 200\,mm beam ROC at one MPC mirror was assumed and the ROC after one pass was predicted by ABCD matrices for different beam sizes. For mode-matching in presence of self-focusing, the beam radius on the mirrors in MPC~1 was reduced by about 30\,\% (intersection blue and black lines) and increased in MPC~2 by circa 15\,\% (intersection red and black lines) in relation to Kerr lens-free mode-matching.  \textbf{c.} ABCD matrix calculations of beam sizes in MPCs 1 and 2 for mode-matching in presence (nonlinear) and absence (linear) of the Kerr effect.}
\label{fig1:Setup}
\end{figure}
\begin{figure}[t]
\centering
\includegraphics[width=.95\linewidth]{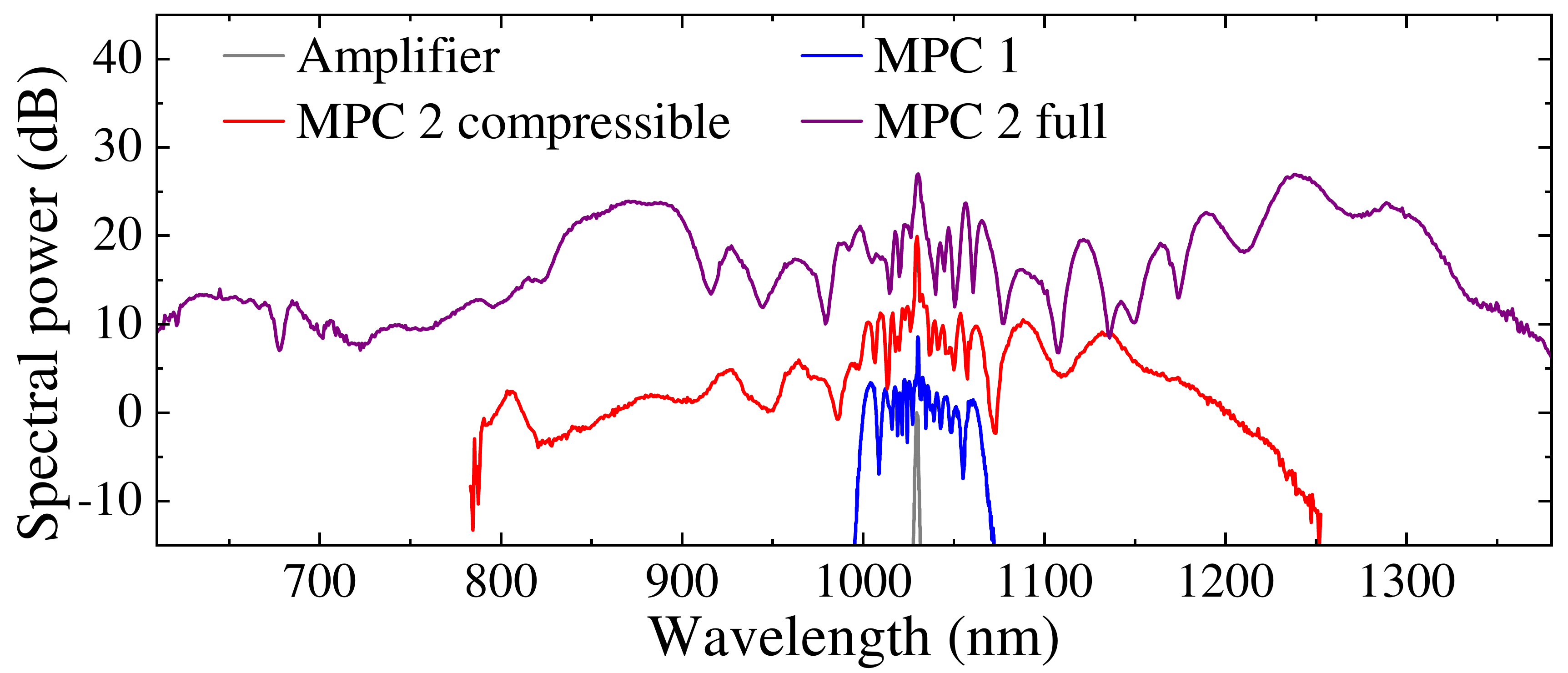}
\caption{The Yb:YAG amplifier spectrum measured with a compact grating spectrometer compared to the broadened spectra after MPCs 1 and 2 which were measured with an optical spectrum analyzer (OSA). The red and the violet lines represent two different MPC 2 settings. The narrower spectrum resulted in the shortest pulses, the broader spectrum covered the full mirror bandwidth. The spectra are offset for the sake of clarity.}
\label{fig2:Spectra}
\end{figure}
\begin{figure}[t]
	\centering
	\includegraphics[width=.94\linewidth]{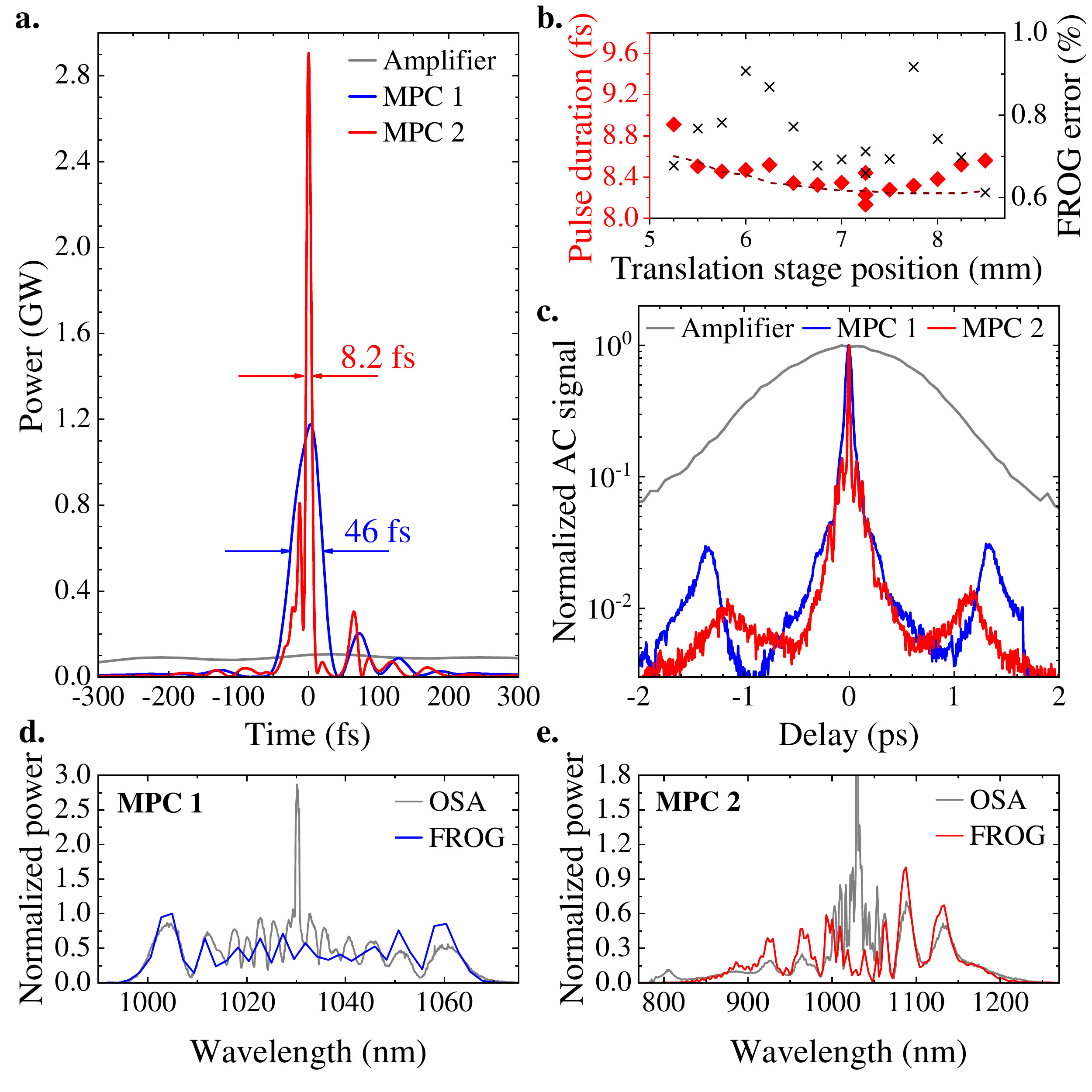}
	\caption{\textbf{a.} Retrieved pulses by FROG from the amplifier and after both compression stages. The 1\,ps long amplifier pulses are only partially shown on the time axis.  \textbf{b.} Retrieved pulse durations (red diamonds) and FROG errors (black crosses) for different amounts of glass in the beam path. A glass wedge with 12$^\circ$ apex angle on a translation stage was moved in 250\,$\mu$m steps, corresponding to approximately 1\,fs$^2$ GDD difference. The dashed line is computed from the electric field of the best retrieved pulse (translation stage position 7.25\,mm) and the theoretical dispersion of the inserted glass. \textbf{c} Autocorrelation (AC) signal extracted directly from the FROG scans. For MPC~2, a step width of 50\,fs was set for the 10\,ps delay range. \textbf{d.}/\textbf{e.} Comparisons between the retrieved spectra after MPC~1 / MPC~2 and the measured OSA spectra. To limit the FROG grid size to 1024$^2$, a delay range of 700\,fs was scanned which explains that the spectral power of the retrieved near-center wavelengths is lower than in the OSA measurement.}
	\label{fig3:pulses}
\end{figure}

The compression setup was based on an Yb:YAG laser and two spectral broadening stages (\Fig{fig1:Setup}a). The laser and the first MPC stage (MPC 1) were similar to the setup reported in ref. \cite{seidel_factor_2022}. The front-end of the amplifier was improved, which led to 15\,\% more pulse energy than in \cite{seidel_factor_2022} and pulses with down to 1\,ps FWHM duration. The amplifier emitted laser bursts every 100\,ms with a variable number of pulses and a 1\,MHz pulse repetition rate. We adjusted the number of pulses to the dynamic range of our measurement devices and typically worked with 150 - 200 pulses per burst. MPC 1 consisted of two quarter-wave stack dielectric mirrors with 200\,mm radius of curvature (ROC) and five 1\,mm thin anti-reflection coated silica substrates. The sixth plate used in ref. \cite{seidel_factor_2022} mainly introduced additional chirp without lowering decisively the 43\,fs Fourier transform limit (FTL) of the MPC~1 output spectrum (\Fig{fig2:Spectra}, blue line). After 68 reflections from chirped mirrors with -200\,fs$^2$ group delay dispersion (GDD), the pulses were compressed to 46\,fs (\Fig{fig3:pulses}a,d). We used input pulses longer than 1\,ps 
%(due to time-constraints, we did not measure the duration) 
to get best compression after MPC 1 at the full input power. This resulted in a pulse energy of 75\muJ\ available for few-cycle pulse compression. The drawback of this configuration was an increase of the $M^2$-parameter from 1.1 to 1.5 after MPC 1 (Table~\ref{tab:M2}), which we related to parasitic four-wave mixing in our previous publication \cite{seidel_factor_2022}. For 1\,ps pulse duration and 96.5\muJ\ energy at the MPC~1 input, we compressed the pulses to 45\,fs while maintaining clearly better $M^2$-values of about 1.3 (Table~\ref{tab:M2}). In this configuration, 61.5\muJ\ pulses could be sent into MPC 2.

%% FIG 2 %%%%%%%%%%%%%%%%%%%%%%%%%%%%%%%%%%%%%%%%%%%%%%%%%%%%%%%%%%%%%%%%%%%%%%%%%%%%%%%%%%%%%%%%%%%%%%%%%%%%
%%%%%%%%%%%%%%%%%%%%%%%%%%%%%%%%%%%%%%%%%%%%%%%%%%%%%%%%%%%%%%%%%%%%%%%%%%%%%%%%%%%%%%%%%%%%%%%%%%
%% TABLE 1 %%
\begin{table}[b!]
	\caption{Results of the $M^2$-measurements.}
	\centering
	\begin{tabular}{ccccc}
		\hline
		& amplifier & MPC 1$^a$ & MPC 1$^b$ & MPC 2$^{a,c}$ \\
		\hline
		$M^2_x$ & 1.16 & 1.43 & 1.28 & 1.45\\
		$M^2_y$ & 1.13 & 1.56 & 1.32 & 1.58 \\
		%2022_07_28_NEPALD_1st_MPC_4pulses_979us_delay_WP1_290_metallic_mirror_no_side_beam
		\hline
	\end{tabular}
	\caption*{$^a$ 128.5\muJ\ at MPC~1, $^b$ 96.5\muJ\ at MPC~1, $^c$ detection up to 1.1\,$\mu$m}
	\label{tab:M2}
\end{table}
%%%%%%%%%%%%%%%%%%%%%%%%%%%%%%%%%%%%%%%%%%%%%%%%%%%%%%%%%%%%%%%%%%%%%%%%%%%%%%%%%%%%%%%%%%%%%%%%%%
%% FIG 3 %%
%% FIG 4 %%
\begin{figure}[t]
	\centering
	\includegraphics[width=.95\linewidth]{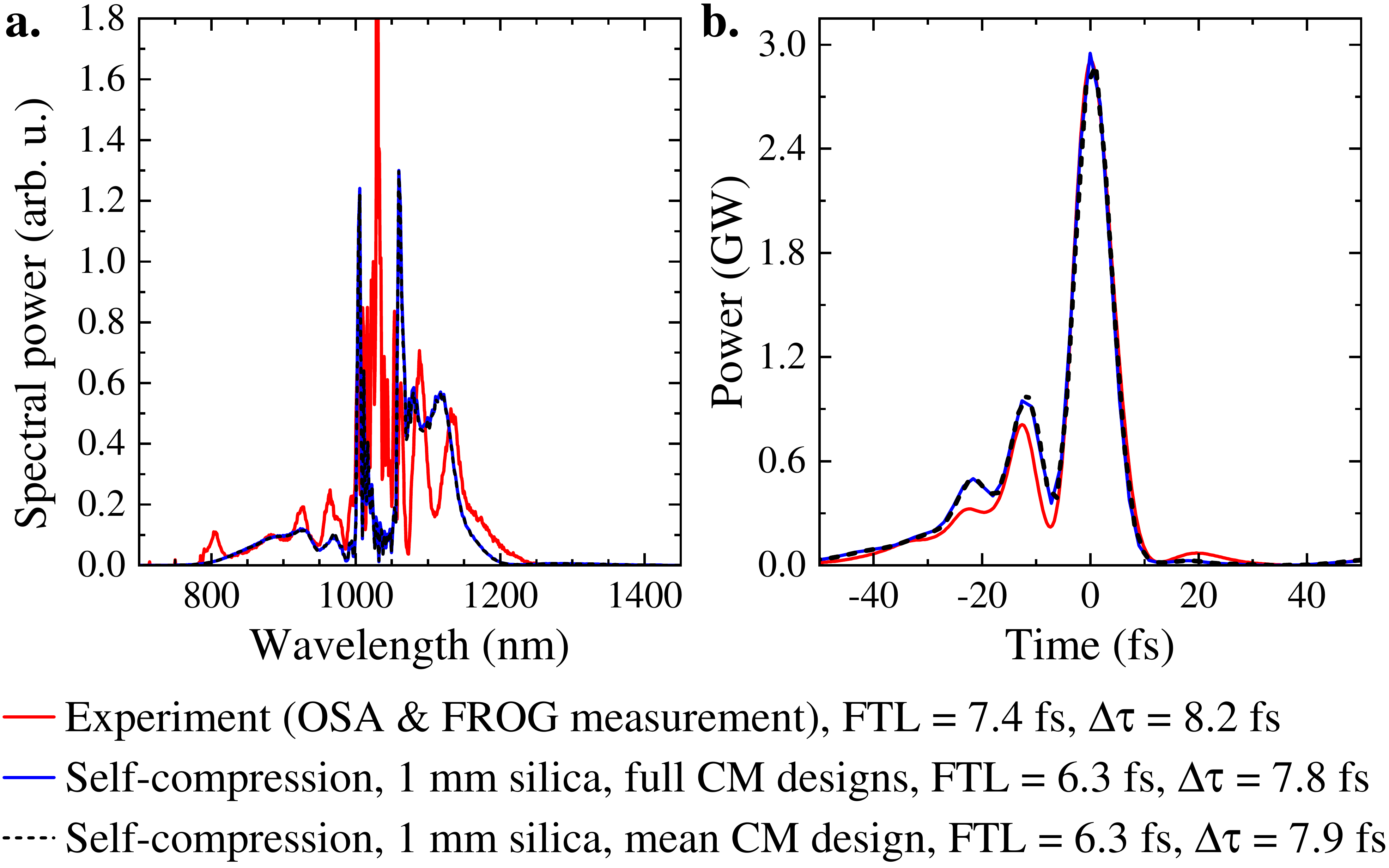}
	\caption{\textbf{a.} Experimental (red line) and simulated (blue/black lines) output spectra of MPC~2. The black dashed line included only averaged properties of the complementary CM pair. \textbf{b.} Corresponding pulses. The pulses plotted with black and blue lines result from self-compression at the end of the seventh MPC roundtrip. $\Delta\tau$ denotes pulse duration.}
	\label{fig4:simulation}
\end{figure}
%%%%%%%%%%%%%%%%%%%%%%%%%%%%%%%%%%%%%%%%%%%%%%%%%%%%%%%%%%%%%%%%%%%%%%%%%%%%%%%%%%%%%%%%%%%%%%%%%%
To accomplish the large compression factors in MPC 1, we used nonlinear mode-matching. That means, we adjusted the distances and refractive powers of the mode-matching lenses under consideration of self-focusing in the nonlinear media \cite{seidel_factor_2022}. The same was done for MPC 2. However, the relative beam size changes with respect to the linear mode-matching setting, which does not account for Kerr lensing, were opposite in both stages (\Fig{fig1:Setup}b,c). In MPC 1, the five silica plates near the cavity center formed a weak waveguide. Therefore, the beam size in the center was larger compared to the linear case. Details are provided in ref.~\cite{seidel_factor_2022}. In contrast, MPC 2 hosted only two silica plates which were located closer to the MPC mirrors than to the beam center. Consequently, the Kerr media merely added to the refractive power of the MPC mirrors causing a smaller beam waist. Nonlinear mode-matching was hence akin to gas-filled MPCs \cite{hanna_nonlinear_2021}. We had to separate the 1\,mm thin silica plates in MPC~2 by about 22\,cm to preserve the compressibility of the pulses. The spectrum measured after 7 roundtrips of the 75\muJ, 46\,fs pulses is plotted in \Fig{fig2:Spectra} (red line). The corresponding 7.4\,fs FTL was enabled by octave-spanning chirped mirrors (CMs, Laseroptik) with 200\,mm ROC, which strongly reduced the net dispersion per pass in MPC~2. To suppress the GDD oscillations inherent to single broadband CMs, an MPC mirror pair with complementary dispersion design was used.\par 
We characterized the compressed pulses by second harmonic frequency-resolved optical gating (FROG) with a 10\,$\mu$m thin BBO crystal cut at $\theta = 29^\circ$. The dispersion-free FROG setup is described in ref.~\cite{seidel_multi-watt_2018}. The shortest pulse duration we retrieved was 8.2\,fs FWHM (\Fig{fig3:pulses}a) corresponding to more than 120 times overall reduction of the pulse duration taking the feasible 1\,ps pulses from the amplifier as reference. A pair of glass wedges (\Fig{fig1:Setup}a) was used to find the best compression point. We compared the retrieved pulse durations from multiple FROG traces at different wedge positions (\Fig{fig3:pulses}b) and obtained very good consistency of the results, such that we infer a $\pm 0.2$\,fs uncertainty of the 8.2\,fs duration. To our knowledge, only bulk-MPCs with at least twice as long pulses were reported before \cite{fritsch_all-solid-state_2018,barbiero_efficient_2021}. We determined a pulse energy of 56\muJ\ after MPC 2. The corresponding 75\,\% transmission of the stage included three bounces off silver mirrors. To minimize the reflection losses of the Kerr media, we placed the silica plates at Brewster's angle into MPC~2. Assuming 97.2\,\% and 99.6\,\% reflectivity of the silver and chirped mirrors, respectively, we deduce an average Fresnel loss of 0.5\,\% per silica-air interface. This shows that polarization rotation due to out-of-plane propagation in the MPC is a minor concern. We attribute this to the tenfold ratio between MPC length and Herriott-pattern diameter. The CM reflectivities were calculated from the broadened spectrum and the mirror design. However, we measured 76.3\,\% transmission of a 12 roundtrip Kerr medium free MPC while we predicted 80.3\,\% transmission from the theoretical reflectivity, implying an average 0.2\,\% difference per pass. Nevertheless, the >99\,\% reflectivity of the CMs is an advantage over (enhanced) silver mirrors, which have been so-far used in all MPCs for sub-10\,fs pulse generation \cite{muller_multipass_2021,rueda_8_2021,daniault_single-stage_2021,hadrich_carrier-envelope_2022}. We note that the CM design exhibits a 0.6\,\% lower reflectivity at 1030\,nm than at the wings of the spectrum after MPC~2. This helps to remove several percent of the residual narrow band radiation emitted by the Yb:YAG amplifier from the compressed pulses. In-fact, the autocorrelation traces of \Fig{fig3:pulses}c show that a side pulse with 1-2\,ps delay from the main peak is suppressed by 5\,dB in comparison to pulses after MPC 1 which is also due to the peak power enhancement of the main pulse. From the pulse energy, the FROG retrieval, which covered a 700\,fs delay range, and the autocorrelation measurement over a 10\,ps range, we derive a peak power of about 2.9\,GW which surpasses the present bulk-MPC record of 2.5\,GW \cite{raab_multi-gigawatt_2022}. Owing to the small net dispersion per pass, we could readily broaden the pulse spectra to fully cover the CM reflectance band from about 0.6\,$\mu$m to 1.4\,$\mu$m by reducing the plate distance and increasing the number of passes in MPC~2. An experiment was conducted with 1\,ps pulses from the laser, 45\,fs, 61.5\muJ\ pulses from MPC 1, 12 roundtrips in MPC~2 and 12\,cm distance between the two Kerr media. This yielded an octave-spanning spectrum with a single-cycle FTL (violet line in \Fig{fig2:Spectra}). However, a FROG measurement showed that it is not possible to compress the pulses close to the spectrum's FTL by the CMs we used. We attribute this to spatio-temporal couplings that arose from increased intensities in the Kerr media. Tailored CMs could compensate for the characteristic bulk-broadening phase \cite{pronin_high-power_2015}. Alternatively, the use of thinner Kerr media like in the multiple plate continuum method promises to push achievable durations in MPC 2 toward the single-cycle regime \cite{seo_high-contrast_2020}.  \par

Figure~\ref{fig4:simulation} compares the experimental results (red lines) with SISYFOS simulations \cite{arisholm_simulation_2012,seidel_factor_2022} of MPC 2. The shortest pulses attainable for two 1\,mm thin silica plates were computed in the course of the seventh roundtrip through MPC~2 omitting the need for post-compression (blue and black lines in \Fig{fig4:simulation}). The net anomalous dispersion was about -10\,fs$^2$ per pass in the simulations. The CM compressor and the glass wedges required in our setup indicate, however, that the experimental net dispersion per pass was closer to 0\,fs$^2$. We attribute the small difference to the imprecise knowledge of the CM mirror dispersion which we did not measure. Nevertheless, the overall agreement between experimental and simulated spectra and pulse shapes is very good. We investigated if the GDD oscillations exhibited by a single CM are detrimental for pulse compression. The blue lines in \Fig{fig4:simulation} show the simulation results under consideration of both complementary mirror designs, whereas the black dashed lines show the results for considering only the averaged reflectivity and GDD of the CM pair. Only minor differences in spectrum and compressed pulse shapes are visible, and thus we conclude that the GDD oscillations of the CMs only marginally influenced the compression results. For the most part, the simulation methods are described in ref.~\cite{seidel_factor_2022}. Owing to shorter input pulses, the Raman response of silica was included in addition to the Kerr effect. 
Reflectivity and GDD used in simulations were blue-shifted from the CM design by 2\,THz owing to slightly lower deposition rates close to the curved mirror edges. 
%Because the deposited layer thickness slightly varied across the curved CMs and the beam in the experiment hit the MPC mirrors close to their edges, the reflectivity and GDD used in simulations were blue-shifted from the CM design by 2\,THz. 
The FROG retrieval from MPC~1 and a fundamental Gaussian were used as pulse and beam shapes, respectively. The simulated pulse energy was set to 33.4\muJ\ in order to match the experimental intensities in the Kerr media. First, the beam area in mode-matched MPCs scales with the M$^2$ factor (here 1.5). Second, Brewster's angle of incidence results in a beam area 45\,\% larger than for normal incidence and a 21\,\% longer optical path through the silica plates which was also taken into account.\par
\begin{figure}[t]
\centering
\begin{minipage}[t]{0.17\linewidth}
	\vspace*{\fill}
	\includegraphics[width=\linewidth]{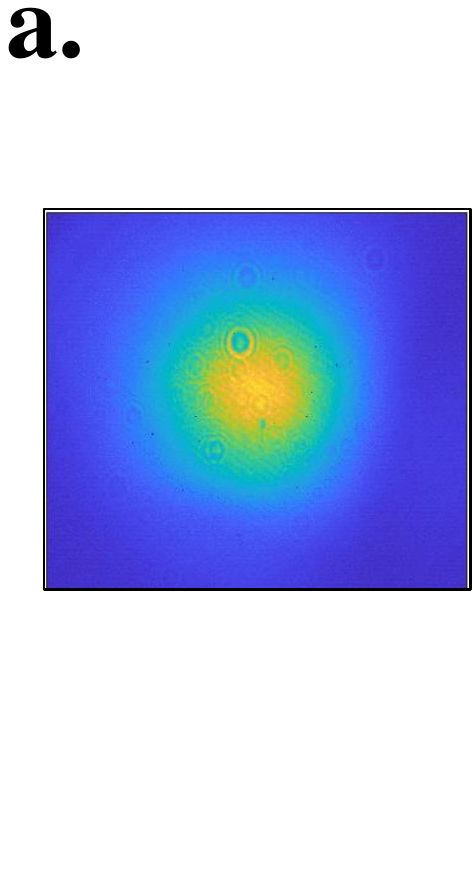}
\end{minipage}
\begin{minipage}[t]{0.78\linewidth}
	\vspace*{-.75mm}
	\includegraphics[width=\linewidth]{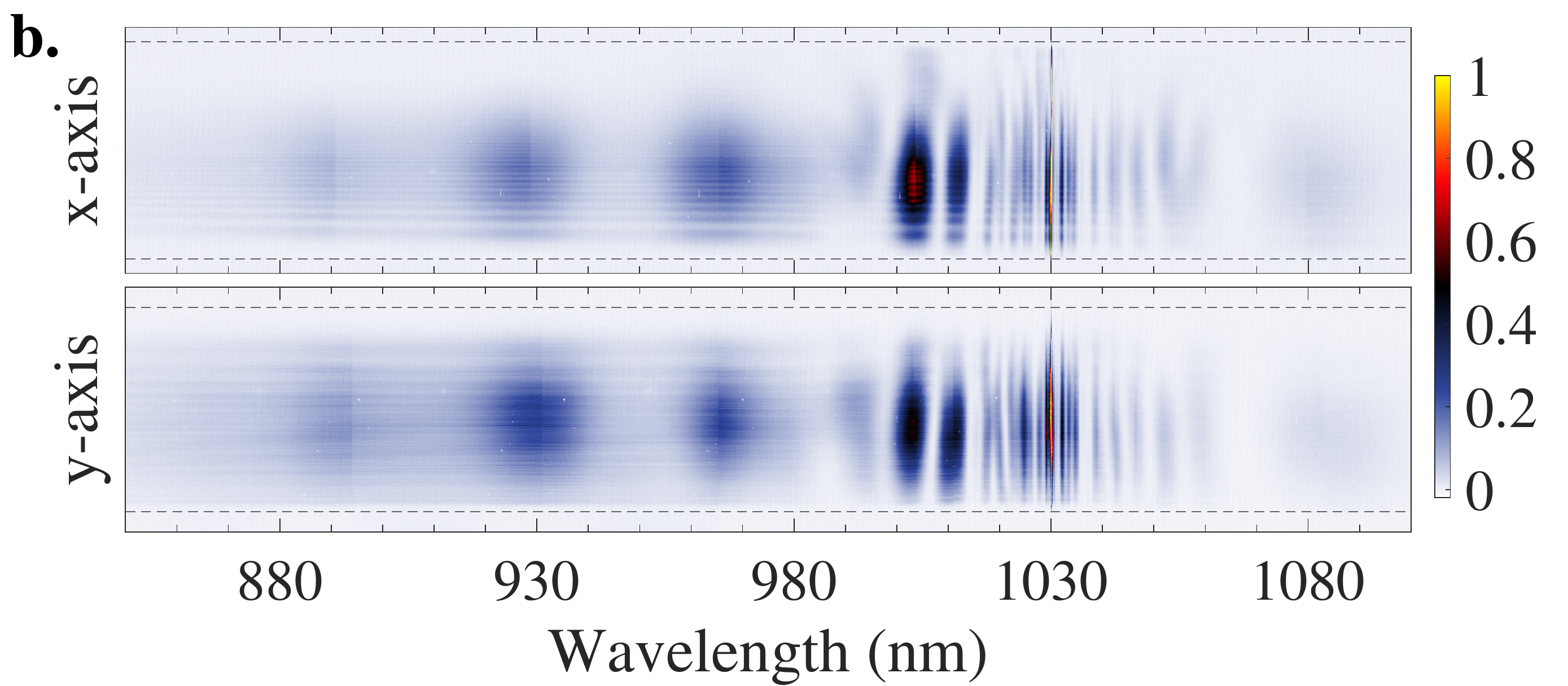}
\end{minipage}
\caption{\textbf{a.} Profile and \textbf{b.} spectral homogeneity of the beam behind MPC 2. The homogeneity was calculated like in ref.~\cite{seidel_ultrafast_2022} over the full width 0.5\,\% maximum of the wavelength integrated power (dashed lines). It was 96.5\,\% along the x- and 97.5\,\% along the y-axis. The spectra had to be recorded over several grating positions of the spectrograph and stitched together in post-processing. The y-axis plot was rotated by 0.43$^\circ$ degree in post-processing of the data. The used Si-based cameras could not respond to wavelengths > 1.1\,$\mu$m.}
\label{fig5:beam}
\end{figure}

We eventually measured the spectral homogeneity of the experimental output beam shown in \Fig{fig5:beam}a with a 4f-imaging spectrograph \cite{seidel_ultrafast_2022,seidel_factor_2022}. Despite Brewster's angle orientation of the Kerr media, the horizontal (x-) and vertical (y-) beam axes exhibited a very good $>96\,\%$ spectral homogeneity of the output beam as usual for MPC compression (\Fig{fig5:beam}b). The determined M$^2$ values were nearly identical to the ones after MPC 1 (Table~\ref{tab:M2}).\par

%% FIG 5 %%
In conclusion, we have turned a ps laser into a few-cycle light source by a sub-m$^2$ footprint two-stage hybrid multi-plate MPC setup that yielded a record-high more than 120-fold pulse duration shortening. The demonstrated multi-GW peak power is well suited for high harmonic generation and probing other strong field phenomena. With better phase control over the attainable octave-spanning spectra and the carrier-envelope offset, a compact MHz rate attosecond source is in reach.\par

\textbf{Acknowledgements.} We thank Cord Arnold (Lund University) and Tobias Gro{\ss} (LASEROPTIK) for fruitful discussions and DESY (Hamburg, Germany), a member of the Helmholtz Association HGF, for the provision of experimental facilities. A.-L. V. acknowledges support from the Swedish Research Council (Vetenskapsrådet grant No. 2019-06275).

%%%%  Bibliography %%%%


\begin{thebibliography}{10}
	\newcommand{\enquote}[1]{``#1''}
	
	\bibitem{brabec_intense_2000}
	T.~Brabec and F.~Krausz, \enquote{Intense few-cycle laser fields: {Frontiers}
		of nonlinear optics,} {{Reviews of Modern Physics}}
	\textbf{72}, 545--591 (2000).
	
	\bibitem{orfanos_attosecond_2019}
	I.~Orfanos, I.~Makos, I.~Liontos, E.~Skantzakis, B.~Förg, D.~Charalambidis,
	and P.~Tzallas, \enquote{Attosecond pulse metrology,}
	{{APL Photonics}} \textbf{4}, 080901 (2019).
	
	\bibitem{ciappina_attosecond_2017}
	M.~F. Ciappina, J.~A. Pérez-Hernández, A.~S. Landsman, W.~A. Okell,
	S.~Zherebtsov, B.~Förg, J.~Schötz, L.~Seiffert, T.~Fennel, T.~Shaaran,
	T.~Zimmermann, A.~Chacón, R.~Guichard, A.~Zaïr, J.~W.~G. Tisch, J.~P.
	Marangos, T.~Witting, A.~Braun, S.~A. Maier, L.~Roso, M.~Krüger,
	P.~Hommelhoff, M.~F. Kling, F.~Krausz, and M.~Lewenstein, \enquote{Attosecond
		physics at the nanoscale,} {{Reports on Progress in
			Physics}} \textbf{80}, 054401 (2017).
	
	\bibitem{mikaelsson_high-repetition_2021}
	S.~Mikaelsson, J.~Vogelsang, C.~Guo, I.~Sytcevich, A.-L. Viotti, F.~Langer,
	Y.-C. Cheng, S.~Nandi, W.~Jin, A.~Olofsson, R.~Weissenbilder, J.~Mauritsson,
	A.~L’Huillier, M.~Gisselbrecht, and C.~L. Arnold, \enquote{A
		high-repetition rate attosecond light source for time-resolved coincidence
		spectroscopy,} {{Nanophotonics}} \textbf{10}, 117--128
	(2021).
	
	\bibitem{kruchinin_colloquium_2018}
	S.~Y. Kruchinin, F.~Krausz, and V.~S. Yakovlev, \enquote{Colloquium:
		{Strong}-field phenomena in periodic systems,} {{Reviews
			of Modern Physics}} \textbf{90}, 021002 (2018).
	
	\bibitem{jimenez-galan_sub-cycle_2021}
	l.~Jiménez-Galán, R.~E.~F. Silva, O.~Smirnova, and M.~Ivanov,
	\enquote{Sub-cycle valleytronics: control of valley polarization using
		few-cycle linearly polarized pulses,} {{Optica}}
	\textbf{8}, 277 (2021).
	
	\bibitem{zuo_highpower_2022}
	J.~Zuo and X.~Lin, \enquote{High‐{Power} {Laser} {Systems},}
	{{Laser \& Photonics Reviews}} \textbf{16}, 2100741
	(2022).
	
	\bibitem{furch_high-power-laser_2022}
	F.~J. Furch, T.~Witting, M.~Osolodkov, F.~Schell, C.~P. Schulz, and M.~J.
	J~Vrakking, \enquote{High power, high repetition rate laser-based sources for
		attosecond science,} {{Journal of Physics: Photonics}}
	\textbf{4}, 032001 (2022).
	
	\bibitem{nagy_high-energy_2021}
	T.~Nagy, P.~Simon, and L.~Veisz, \enquote{High-energy few-cycle pulses:
		post-compression techniques,} {{Advances in Physics: X}}
	\textbf{6}, 1845795 (2021).
	
	\bibitem{schulte_nonlinear_2016}
	J.~Schulte, T.~Sartorius, J.~Weitenberg, A.~Vernaleken, and P.~Russbueldt,
	\enquote{Nonlinear pulse compression in a multi-pass cell,}
	{{Optics Letters}} \textbf{41}, 4511 (2016).
	
	\bibitem{viotti_multi-pass_2022}
	A.-L. Viotti, M.~Seidel, E.~Escoto, S.~Rajhans, W.~P. Leemans, I.~Hartl, and
	C.~M. Heyl, \enquote{Multi-pass cells for post-compression of ultrashort
		laser pulses,} {{Optica}} \textbf{9}, 197 (2022).
	
	\bibitem{hanna_MPC_rev_2021}
	M.~Hanna, F.~Guichard, N.~Daher, Q.~Bournet, X.~Délen, and P.~Georges,
	\enquote{Nonlinear {Optics} in {Multipass} {Cells},}
	{{Laser \& Photonics Reviews}} \textbf{15}, 2100220
	(2021).
	
	\bibitem{balla_postcompression_2020}
	P.~Balla, A.~Bin~Wahid, I.~Sytcevich, C.~Guo, A.-L. Viotti, L.~Silletti,
	A.~Cartella, S.~Alisauskas, H.~Tavakol, U.~Grosse-Wortmann, A.~Schönberg,
	M.~Seidel, A.~Trabattoni, B.~Manschwetus, T.~Lang, F.~Calegari, A.~Couairon,
	A.~L’Huillier, C.~L. Arnold, I.~Hartl, and C.~M. Heyl,
	\enquote{Postcompression of picosecond pulses into the few-cycle regime,}
	{{Optics Letters}} \textbf{45}, 2572 (2020).
	
	\bibitem{muller_multipass_2021}
	M.~Müller, J.~Buldt, H.~Stark, C.~Grebing, and J.~Limpert, \enquote{Multipass
		cell for high-power few-cycle compression,} {{Optics
			Letters}} \textbf{46}, 2678 (2021).
	
	\bibitem{rueda_8_2021}
	P.~Rueda, F.~Videla, T.~Witting, G.~A. Torchia, and F.~J. Furch, \enquote{8 fs
		laser pulses from a compact gas-filled multi-pass cell,}
	{{Optics Express}} \textbf{29}, 27004--27013 (2021).
	
	\bibitem{daniault_single-stage_2021}
	L.~Daniault, Z.~Cheng, J.~Kaur, J.-F. Hergott, F.~R\'{e}au, O.~Tcherbakoff,
	N.~Daher, X.~D\'{e}len, M.~Hanna, and R.~Lopez-Martens, \enquote{Single-stage
		few-cycle nonlinear compression of millijoule energy {Ti:Sa} femtosecond
		pulses in a multipass cell,} {{Opt. Lett.}} \textbf{46},
	5264--5267 (2021).
	
	\bibitem{hadrich_carrier-envelope_2022}
	S.~Hädrich, E.~Shestaev, M.~Tschernajew, F.~Stutzki, N.~Walther, F.~Just,
	M.~Kienel, I.~Seres, P.~Jójárt, Z.~Bengery, B.~Gilicze, Z.~Várallyay,
	A.~Börzsönyi, M.~Müller, C.~Grebing, A.~Klenke, D.~Hoff, G.~G. Paulus,
	T.~Eidam, and J.~Limpert, \enquote{Carrier-envelope phase stable few-cycle
		laser system delivering more than 100 {W}, 1 {mJ}, sub-2-cycle pulses,}
	{{Optics Letters}} \textbf{47}, 1537 (2022).
	
	\bibitem{lu_generation_2014}
	C.-H. Lu, Y.-J. Tsou, H.-Y. Chen, B.-H. Chen, Y.-C. Cheng, S.-D. Yang, M.-C.
	Chen, C.-C. Hsu, and A.~H. Kung, \enquote{Generation of intense
		supercontinuum in condensed media,} {{Optica}}
	\textbf{1}, 400 (2014).
	
	\bibitem{lu_greater_2019}
	C.-H. Lu, W.-H. Wu, S.-H. Kuo, J.-Y. Guo, M.-C. Chen, S.-D. Yang, and A.~H.
	Kung, \enquote{Greater than 50 times compression of 1030 nm {Yb}:{KGW} laser
		pulses to single-cycle duration,} {{Optics Express}}
	\textbf{27}, 15638 (2019).
	
	\bibitem{seo_high-contrast_2020}
	M.~Seo, K.~Tsendsuren, S.~Mitra, M.~Kling, and D.~Kim, \enquote{High-contrast,
		intense single-cycle pulses from an all thin-solid-plate setup,}
	{{Optics Letters}} \textbf{45}, 367 (2020).
	
	\bibitem{seidel_factor_2022}
	M.~Seidel, P.~Balla, C.~Li, G.~Arisholm, L.~Winkelmann, I.~Hartl, and C.~M.
	Heyl, \enquote{Factor 30 {Pulse} {Compression} by {Hybrid} {Multipass}
		{Multiplate} {Spectral} {Broadening},} {{Ultrafast
			Science}} \textbf{2022}, 9754919 (2022).
	
	\bibitem{seidel_ultrafast_2022}
	M.~Seidel, F.~Pressacco, O.~Akcaalan, T.~Binhammer, J.~Darvill, N.~Ekanayake,
	M.~Frede, U.~Grosse‐Wortmann, M.~Heber, C.~M. Heyl, D.~Kutnyakhov, C.~Li,
	C.~Mohr, J.~Müller, O.~Puncken, H.~Redlin, N.~Schirmel, S.~Schulz,
	A.~Swiderski, H.~Tavakol, H.~Tünnermann, C.~Vidoli, L.~Wenthaus, N.~Wind,
	L.~Winkelmann, B.~Manschwetus, and I.~Hartl, \enquote{Ultrafast
		{MHz}‐{Rate} {Burst}‐{Mode} {Pump}–{Probe} {Laser} for the {FLASH}
		{FEL} {Facility} {Based} on {Nonlinear} {Compression} of ps‐{Level}
		{Pulses} from an {Yb}‐{Amplifier} {Chain},} {{Laser \&
			Photonics Reviews}} \textbf{16}, 2100268 (2022).
	
	\bibitem{hanna_nonlinear_2021}
	M.~Hanna, L.~Daniault, F.~Guichard, N.~Daher, X.~Délen, R.~Lopez-Martens, and
	P.~Georges, \enquote{Nonlinear beam matching to gas-filled multipass cells,}
	{{OSA Continuum}} \textbf{4}, 732 (2021).
	
	\bibitem{seidel_multi-watt_2018}
	M.~Seidel, X.~Xiao, S.~A. Hussain, G.~Arisholm, A.~Hartung, K.~T. Zawilski,
	P.~G. Schunemann, F.~Habel, M.~Trubetskov, V.~Pervak, O.~Pronin, and
	F.~Krausz, \enquote{Multi-watt, multi-octave, mid-infrared femtosecond
		source,} {{Science Advances}} \textbf{4}, eaaq1526
	(2018).
	
	\bibitem{fritsch_all-solid-state_2018}
	K.~Fritsch, M.~Poetzlberger, V.~Pervak, J.~Brons, and O.~Pronin,
	\enquote{All-solid-state multipass spectral broadening to sub-20 fs,}
	{{Optics Letters}} \textbf{43}, 4643 (2018).
	
	\bibitem{barbiero_efficient_2021}
	G.~Barbiero, H.~Wang, M.~Graßl, S.~Gröbmeyer, D.~Kimbaras, M.~Neuhaus,
	V.~Pervak, T.~Nubbemeyer, H.~Fattahi, and M.~F. Kling, \enquote{Efficient
		nonlinear compression of a thin-disk oscillator to 8.5 fs at 55 {W} average
		power,} {{Optics Letters}} \textbf{46}, 5304 (2021).
	
	\bibitem{raab_multi-gigawatt_2022}
	A.-K. Raab, M.~Seidel, C.~Guo, I.~Sytcevich, G.~Arisholm, A.~L'Huillier, C.~L.
	Arnold, and A.-L. Viotti, \enquote{Multi-gigawatt peak power post-compression
		in a bulk multi-pass cell at a high repetition rate,}
	{{Opt. Lett.}} \textbf{47}, 5084--5087 (2022).
	
	\bibitem{pronin_high-power_2015}
	O.~Pronin, M.~Seidel, F.~Lücking, J.~Brons, E.~Fedulova, M.~Trubetskov,
	V.~Pervak, A.~Apolonski, T.~Udem, and F.~Krausz, \enquote{High-power
		multi-megahertz source of waveform-stabilized few-cycle light,}
	{{Nature Communications}} \textbf{6}, 6988 (2015).
	
	\bibitem{arisholm_simulation_2012}
	G.~Arisholm and H.~Fonnum, \emph{Simulation {System} {For} {Optical} {Science}
		({SISYFOS}) – tutorial,version 2}, vol. 21/01183 of \emph{{FFI}-rapport}
	(Norwegian Defence Research Establishment (FFI), 2021).
	
\end{thebibliography}
\end{document}